\documentclass[prd,preprint,superscriptaddress,amsmath,amssymb,nofootinbib]{revtex4}
\pdfoutput=1

\usepackage{graphicx,color,slashed}% Include figure files

\def\be{\begin{eqnarray}}
\def\ed{\end{eqnarray}}

\newcommand{\mytabnote}{\footnote{\label{tab} We follow the formulae in \cite{Nobes:2000pm} to estimate the form factor values.}}

\begin{document}

%{\begin{flushright}{xxx}
%\end{flushright}}

\title{\bf \Large Constraint on the branching ratio of $B^{-}_c\to \tau\bar\nu$ 
from LEP1 and consequences for $R(D^{(*)})$ anomaly}

\author{A.G. Akeroyd}
\email{a.g.akeroyd@soton.ac.uk}
\affiliation{School of Physics and Astronomy, University of Southampton,
Highfield, Southampton SO17 1BJ, United Kingdom}

\author{Chuan-Hung Chen}
\email{physchen@mail.ncku.edu.tw}
\affiliation{Department of Physics, National Cheng-Kung University, Tainan 70101, Taiwan}

\date{\today}% It is always \today, today,

\begin{abstract}
Recently there has been interest in the correlation between
$R(D^*)$ and the branching ratio (BR) of $B^-_c\to \tau\bar\nu$ in models 
with a charged scalar $H^\pm$. Any enhancement of $R(D^*)$ by $H^\pm$ alone
(in order to agree with current data) also enhances BR($B^-_c\to \tau\bar\nu$), 
for which there has been no direct search at hadron colliders. 
We show that LEP data taken at the $Z$ peak requires
BR($B^-_c\to \tau\bar\nu)\lesssim 10\%$, and  
this constraint is significantly stronger than 
the recent constraint BR($B^-_c\to \tau\bar\nu) \lesssim 30\%$ 
from considering the lifetime of $B_c$. 
In order to respect this new constraint, any explanation of the 
$R(D)$ and $R(D^*)$ anomaly in terms of $H^\pm$ 
alone would require the future measurements of $R(D^*)$ 
to be even closer to the Standard Model prediction. A stronger limit
on BR$(B^-_c\to \tau\bar\nu)$ (or its first measurement) would
be obtained if the L3 collaboration 
used all its data taken at the $Z$ peak.

\end{abstract}

\maketitle

\section{Introduction}

The $B_c$ meson is the ground state of a quarkonium system that is composed of a $c$ and a $b$ quark. Prior to the operation of the LHC 
there were only a few measurements of its properties from Tevatron data 
\cite{Abe:1998fb,Abulencia:2005usa,Abulencia:2006zu,Aaltonen:2007gv}.
The LHC experiments (in particular LHCb) promise the first detailed study of $B_c$. More precise measurements
of its mass and lifetime are now available, and several decay channels have been observed for the first time. It is well 
known that precise measurements of the branching ratios (BRs) of hadrons play an important role in constraining 
the properties of new physics particles. The measured BRs of decays such as 
$b \to s \gamma$, $B^-_u \to \tau \bar\nu$ and $B^-_u \to D^{(*)} \tau \bar\nu$ 
all provide constraints on the coupling constants and the masses of new physics particles, and often such constraints  
are stronger than those that are derived from direct searches at the LHC.
There have been a few works on the potential of the $B_c$ meson to probe the presence of new physics particles.  
In particular the BR of the leptonic decay 
$B^-_c\to \tau \bar\nu$ could be significantly enhanced by a charged Higgs 
boson ($H^\pm$) \cite{Du:1997pm,Mangano:1997md,Akeroyd:2008ac}
or by supersymmetric particles with specific R-parity violating couplings \cite{Baek:1999ch,Akeroyd:2002cs}.

The potential of the $B_c$ meson to constrain the properties of 
new physics particles has attracted renewed 
attention recently. It was shown in \cite{Alonso:2016oyd} 
that the measured value of the lifetime of 
$B_c$ disfavours an explanation of 
the $R(D)$ and $R(D^*)$ anomaly (in $B^-_u \to D^{(*)} \tau\bar \nu$ decays) 
in terms of an $H^\pm$ alone\footnote{For a study of the impact of the
$B_c$ lifetime on a leptoquark explanation of the $R(D)$ and $R(D^*)$ anomaly 
see \cite{Li:2016vvp}.}. This is because any enhancement
of $R(D^*)$ by an $H^\pm$ would also cause an enhancement of the
BR of the unobserved decay $B^-_c\to \tau\bar\nu$. 
In order to comply with the current world average of 
the $B_c$ lifetime it was shown that BR$(B^-_c \to \tau \bar \nu)\lesssim 30\%$
is necessary, but accommodating the measured values 
of $R(D)$ and $R(D^*)$ by $H^\pm$ alone
would require BR$(B^-_c \to \tau \bar \nu)> 30\%$. 
%Hence it is concluded in \cite{Alonso:2016oyd}  that an explanation of 
%the $R(D%),R(D^*)$
% anomaly from $H^\pm$ alone is disfavoured at present. 

In this paper we derive a stronger bound on BR$(B^-_c \to \tau \bar \nu)$ than that obtained from the lifetime of $B_c$. 
LEP data taken at the $Z$ peak constrained a combination of $B^-_u \to \tau \bar \nu$ and $B^-_c \to \tau \bar \nu$
\cite{Acciarri:1996bv,Abreu:1999xe,Barate:2000rc}. This was first
pointed out in \cite{Mangano:1997md}, and in an earlier work 
\cite{Akeroyd:2008ac} 
we showed that a signal for the sum of the processes $B^-_u \to \tau \bar \nu$ 
and $B^-_c \to \tau \bar\nu$ might be observed if the 
L3 collaboration 
(which had the strongest limits \cite{Acciarri:1996bv} 
from the LEP collaborations) performed the search with their full data sample.
 A crucial input parameter for the detection prospects of $B^-_c \to 
\tau \bar\nu$ is the transition probability 
(denoted by $f_c$) of a $b$ quark hadronising to a $B_c$. In \cite{Akeroyd:2008ac} the value of $f_c$  
was obtained (with sizeable errors) from early Tevatron measurements. 

%The explicit bound on BR$(B_c \to \tau \bar\nu)$ from using the L3 limit was not given. 
Building on the analysis of \cite{Akeroyd:2008ac}, we first obtain a much more precise evaluation of $f_c$ from 
measurements of $B_c$ production/decay with the full Tevatron data \cite{Aaltonen:2016dra} and from LHC measurements 
\cite{Khachatryan:2014nfa,Aaij:2012dd,Aaij:2014ija,Aaij:2014jxa}.
We then derive a formula for the bound on BR$(B^-_c \to \tau \bar \nu)$ from LEP data, 
which was not obtained in \cite{Akeroyd:2008ac}. 
The bound can be expressed in terms of experimentally
determined quantities and just one theoretical input parameter, which is 
the BR of $B^-_c\to J/\psi \ell \bar\nu$.
Guided by recent lattice QCD calculations of the form factors for 
$B^-_c\to J/\psi \ell \bar\nu$, we present the
preferred range for its theoretical BR. We then obtain a bound on BR$(B^-_c \to \tau \bar \nu)$
that
is considerably stronger than the bound in \cite{Alonso:2016oyd} from considering the lifetime of $B_c$. 
Finally we discuss the consequences of this stronger bound on  BR$(B^-_c \to \tau \bar \nu)$
for an interpretation of the $R(D)$ and $R(D^*)$ anomaly in terms of 
an $H^\pm$ alone.

%Our work is organised as follows. In section II we present a derivation the bou%nd on BR$(B_c \to \tau \bar \nu)$, and in
%section III we consider its impact on  the $R(D)$ and $R(D^*)$ anomaly. Secton %IV contains our conclusions.

\section{The decay $B^-_c\to \tau\bar\nu$ and searches at LEP}

The LEP searches for $B^-_u\to \tau\bar\nu$ 
with data taken at $\sqrt s\sim 91$ GeV (the ``$Z$ peak'')
\cite{Acciarri:1996bv,Abreu:1999xe,Barate:2000rc} 
were sensitive to $\tau\bar\nu$ events originating
from both $B^-_u\to \tau\bar\nu$ and $B^-_c\to \tau\bar\nu$ \cite{Mangano:1997md}.
Hence the published limits constrain an
``effective branching ratio'' defined by:
\begin{equation}
{\rm BR}_{\rm eff}={\rm BR}(B^-_u\to \tau \bar\nu)\left(1+\frac{N_c}
 {N_u}\right) \,.
 \label{BReff}
\end{equation}
This expression applies to all searches for $B^-_u\to \tau\bar\nu$ at 
$e^+e^-$ colliders with data taken at the
$Z$ peak. For searches at the $\Upsilon(4S)$ (i.e. the BABAR and BELLE experiments operating with
$\sqrt s\sim 10.6$ GeV) the $B_c$ meson cannot be produced. Thus in those experiments $N_c=0$ 
and ${\rm BR}_{\rm eff}$=BR($B^-_u\to \tau\bar\nu$). At the $Z$ peak one has the following expression for $N_c/N_u$: 
\begin{equation}
\frac{N_c}{N_u}=\frac{f_c}{f_u}
\frac{{\rm BR}(B^-_c\to \tau\bar\nu)}{{\rm BR} (B^-_u\to \tau\bar\nu)}\,. 
\end{equation}
Substituting for $N_c/N_u$ in eq.~(\ref{BReff}) 
gives rise to following expression for 
${\rm BR}(B^-_c\to \tau\bar\nu)$ in terms of ${\rm BR}_{\rm eff}$:
\begin{equation}
{\rm BR}(B^-_c\to \tau\bar\nu)=\frac{f_u}{f_c}\left[{\rm BR}_{\rm eff}-{\rm BR} 
(B^-_u\to \tau\bar\nu)\right]\,.
\label{Bc}
\end{equation}
%The measured values of $f_u$ \cite{Amhis:2016xyh} are given in 
%Table~\ref{Fu}. 
%\begin{table}
%\center{\begin{tabular}{|l|ccc|}
%\hline
%& LEP & Tevatron & Average \\\hline
%$f_u$ & $0.407\pm 0.007$ & $0.344\pm 0.021$ & $0.404 \pm 0.006$ \\
%\hline
%\end{tabular} }
%\caption{Measured values of $f_u$ at LEP and Tevatron, and their average.}
%\label{Fu}
%\end{table}
Here ${\rm BR}(B^-_u\to \tau\bar\nu)= \left(1.06 \pm 0.19 \right)\times 10^{-4}\,$, which
is the world average \cite{Amhis:2016xyh} of BABAR and BELLE measurements.
%\begin{equation}
%{\rm BR}(B_u\to \tau\bar\nu)= \left(1.06 \pm 0.19 \right)
%\times 10^{-4}\,.
%\label{Btaunu_exp}
%\end{equation}
The L3 collaboration obtained the bound ${\rm BR_{eff}} 
<5.7\times 10^{-4}$ \cite{Acciarri:1996bv}.
If $f_c/f_u$ is known then a bound on ${\rm BR}(B^-_c\to \tau\bar\nu)$
can be derived from eq.~(\ref{Bc}). The value of
$f_c/f_u$ can be obtained from Tevatron and LHC data (see later).

In the Tevatron Run I and II the following ratio was measured:
\begin{equation}
{\cal R_{\ell}} = \frac{\sigma(B_c)\cdot {\rm BR}(B^-_c\to J/\psi
\ell \bar\nu)}
   {\sigma(B_u)\cdot {\rm BR}(B^-_u\to J/\psi K^-)}\,.
\label{tevatron98}
 \end{equation}
Tevatron Run I data with 0.11 fb$^{-1}$ gave the result 
${\cal R_{\ell}}=0.13\pm 0.06$ \cite{Abe:1998fb}.
Tevatron Run II data with 0.36 fb$^{-1}$ gave ${\cal R_{\ell}}=0.28\pm 0.07$
\cite{Abulencia:2006zu}, and this measurement was used in the analysis of \cite{Akeroyd:2008ac}
when extracting $f_c/f_u$.
Recently, using the full CDF Run II data 
(8.7 fb$^{-1}$) the result ${\cal R_{\ell}}=0.211\pm 0.012\pm 0.021$ was obtained \cite{Aaltonen:2016dra}.
The transition probability
$f_c$ determines $\sigma(B_c)$ and several 
theoretical calculations are available for BR($B^-_c\to J/\psi
\ell \bar \nu)$ \cite{Chang:1992pt,Anisimov:1998uk,Kiselev:1999sc,AbdElHady:1999xh,Colangelo:1999zn,Nobes:2000pm,Ebert:2003cn,Ivanov:2005fd,Ivanov:2006ni,Hernandez:2006gt,Wang:2008xt,Ke:2013yka}. 

\begin{figure}[phtb]
\includegraphics[width=85mm]{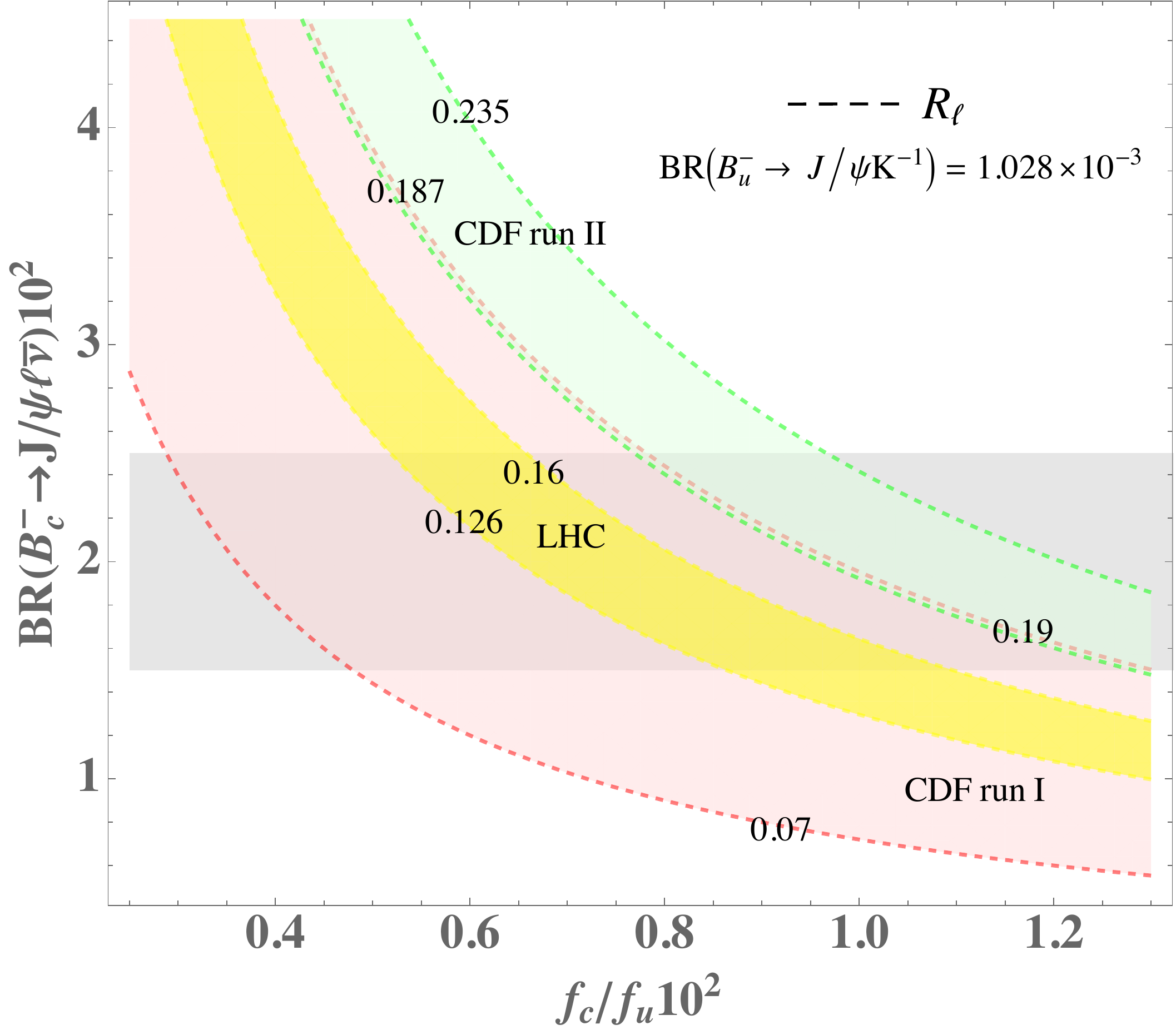}
 \caption{$R_{\ell}$ as a function of $BR(B^-_c \to J/\psi \ell \bar\nu)$ and 
$f_c/f_u$, where the different bands denote the results from the CDF Run I (red), Run II (green), and LHC (yellow) with $1\sigma$ errors. }
\label{fig:R_l}
\end{figure}

The LHC collaborations have not yet measured $R_\ell$ directly. However,
two related ratios have been measured, from which a measurement of 
$R_\ell$ can be obtained.
The ratio ${{\cal R}_{\pi/K}}$ is defined as:
\begin{equation}
{{\cal R}_{\pi/K}} = \frac{\sigma(B_c)\cdot {\rm BR}(B^-_c\to J/\psi \pi^-)}
   {\sigma(B_u)\cdot {\rm BR}(B^-_u\to J/\psi K^-)}\,.
\label{R_piK}
 \end{equation}
The measurements at CMS with $\sqrt s=7$ TeV and 5 ${\rm fb}^{-1}$
\cite{Khachatryan:2014nfa}, LHCb collaboration with $\sqrt s=7$ TeV and
0.37 ${\rm fb}^{-1}$
\cite{Aaij:2012dd}, and LHCb collaboration with $\sqrt s=8$ TeV and
2 ${\rm fb}^{-1}$
\cite{Aaij:2014ija} have been averaged in \cite{Amhis:2016xyh}, with the
result ${{\cal R}_{\pi/K}}=(6.72\pm 0.19)\times 10^{-3}$. 
The ratio ${{\cal R}_{\pi/\mu}}$ is defined as:
\begin{equation}
{{\cal R}_{\pi/\mu}} = \frac{ {\rm BR}(B^-_c\to J/\psi 
\pi^-)}
   {\ {\rm BR}(B^-_c \to J/\psi \mu \bar\nu)}\,.
\label{R_pimu}
 \end{equation}
The measured value at LHCb with $\sqrt s=7$ TeV and 1 ${\rm fb}^{-1}$  
is ${{\cal R}_{\pi/\mu}}=0.0469\pm 0.0054$ \cite{Aaij:2014jxa}. 
Now the ratio ${\cal R}_{\pi/K}$ in eq.~(\ref{R_piK}) can be written 
as:
\begin{equation}
{\cal R}_{\pi/K}=
%\frac{\sigma(B_c)}{\sigma(B_u)}
%\frac{{\rm BR}(B_c\to J/\Psi \mu\nu_\mu)}{{\rm BR}(B_u\to J/\Psi K^-)}
%\frac{{\rm BR}(B_u\to J/\Psi \pi^-)}{{\rm BR}(B_c\to J/\Psi \mu\nu)}=
{\cal R}_{\ell} \cdot {\cal R}_{\pi/\mu}\,. 
\end{equation}
Hence ${\cal R}_{\ell}$ can be extracted from the LHCb measurements
of ${\cal R}_{\pi/\mu}$ and ${\cal R}_{\pi/K}$. One obtains
\begin{equation}
{\cal R}_{\ell} = \frac{{\cal R}_{\pi/K}}{{\cal R}_{\pi/\mu}}
%%=\frac{(6.72\pm 0.19)\times 10^{-3}}{0.0469\pm 0.0054}
 =0.143\pm 0.017\,.
\end{equation}
\begin{table}
\center{\begin{tabular}{|l|c|c|c|c|}
\hline
& Tevatron Run I & Tevatron Run II & Average I+II & LHC \\\hline
${\cal R}_{\ell}$ & $0.13\pm 0.06$ & $0.211\pm 0.024$ & $0.171\pm 0.032$
& $0.143 \pm 0.017$ \\
\hline
\end{tabular} }
\caption{Measured values of ${\cal R}_{\ell}$ 
at Tevatron Run I and II, average of Run I+II, and LHC.}
\label{Rl_measure}
\end{table}
Since $\sigma(B_c)/\sigma(B_u)=f_c/f_u$ then from the definition of 
${\cal R}_{\ell}$ one has:
%can also be written as
%\begin{equation}
%{\cal R}_{\ell} = \frac{f_c\cdot {\rm BR}(B_c\to J/\psi
%\ell^\pm \nu_{\ell})}
%   {f_u\cdot {\rm BR}(B\to J/\psi K^+)}\,.
%\end{equation}
%Therefore one has
\begin{equation}
\frac{f_c}{f_u}=\frac{{\rm BR}(B^-_u\to J/\psi K^-)}
{{\rm BR}(B^-_c\to J/\psi\ell \bar\nu)}{\cal R}_{\ell}\,.
\end{equation}
Here ${\rm BR}(B^-_u\to J/\psi K^-)=(1.028\pm 0.04)\times 10^{-3}$. Using 
the measured values of ${\cal R}_{\ell}$ from the Tevatron and LHC gives the
following expression: 
\begin{equation}
%\frac{f_c}{f_u}=\frac{(1.758\pm 0.336)\times 10^{-4}}{{\rm BR}(B_c\to J/\psi
%\ell^\pm \nu_{\ell})}\,\,\,\,\,\,{\rm (Tevatron\,data)}\,.
\frac{f_c}{f_u}=\frac{10^{-4}}{{\rm BR}(B^-_c\to J/\psi \ell \bar \nu)} 
\left\{\begin{array} {c}
    1.758 \pm 0.336 ~~~ {\rm (Tevatron\,data)}\,, \\ 
    1.470 \pm 0.184 ~~~ {\rm (LHC\,data)} \,.~~~~~ \end{array}\right.
\label{fcfu_tev_lhc}
\end{equation}
%Using the measured value of ${\cal R}_{\ell}$ from the LHC gives the
%following expression: 
%\begin{equation}
%\frac{f_c}{f_u}=\frac{(1.470\pm 0.184)\times 10^{-4}}{{\rm BR}(B_c\to J/\psi
%\ell^\pm \nu_{\ell})}\,\,\,\,\,\,{\rm (LHC\,data)}\,.
%\label{fcfu_lhc}
%\end{equation}
In Fig.~\ref{fig:R_l} we display contours of ${\cal
R}_{\ell}$ as a function of ${\rm BR}(B^-_c\to J/\psi \ell \bar \nu)$ and
$f_c/f_u$, and the band denotes the
prediction of the various theoretical calculations
for ${\rm BR}(B^-_c\to J/\psi \ell \bar \nu)$ whose values lie in the range 
$(1.5\sim 2.5)\%$
\cite{Chang:1992pt,Anisimov:1998uk,Kiselev:1999sc,AbdElHady:1999xh,Colangelo:1999zn,Nobes:2000pm,Ebert:2003cn,Ivanov:2005fd,Ivanov:2006ni,Hernandez:2006gt,Wang:2008xt,Ke:2013yka}.
%We fix $f_u=0.404$. Clearly one has $1\times 10^{-3} < f_c < 5\times 10^{-3}$, and
%this result is an update of our result for $f_c$ in \cite{Akeroyd:2008ac}.

%%%%

\begin{figure}[phtb]
\includegraphics[width=85mm]{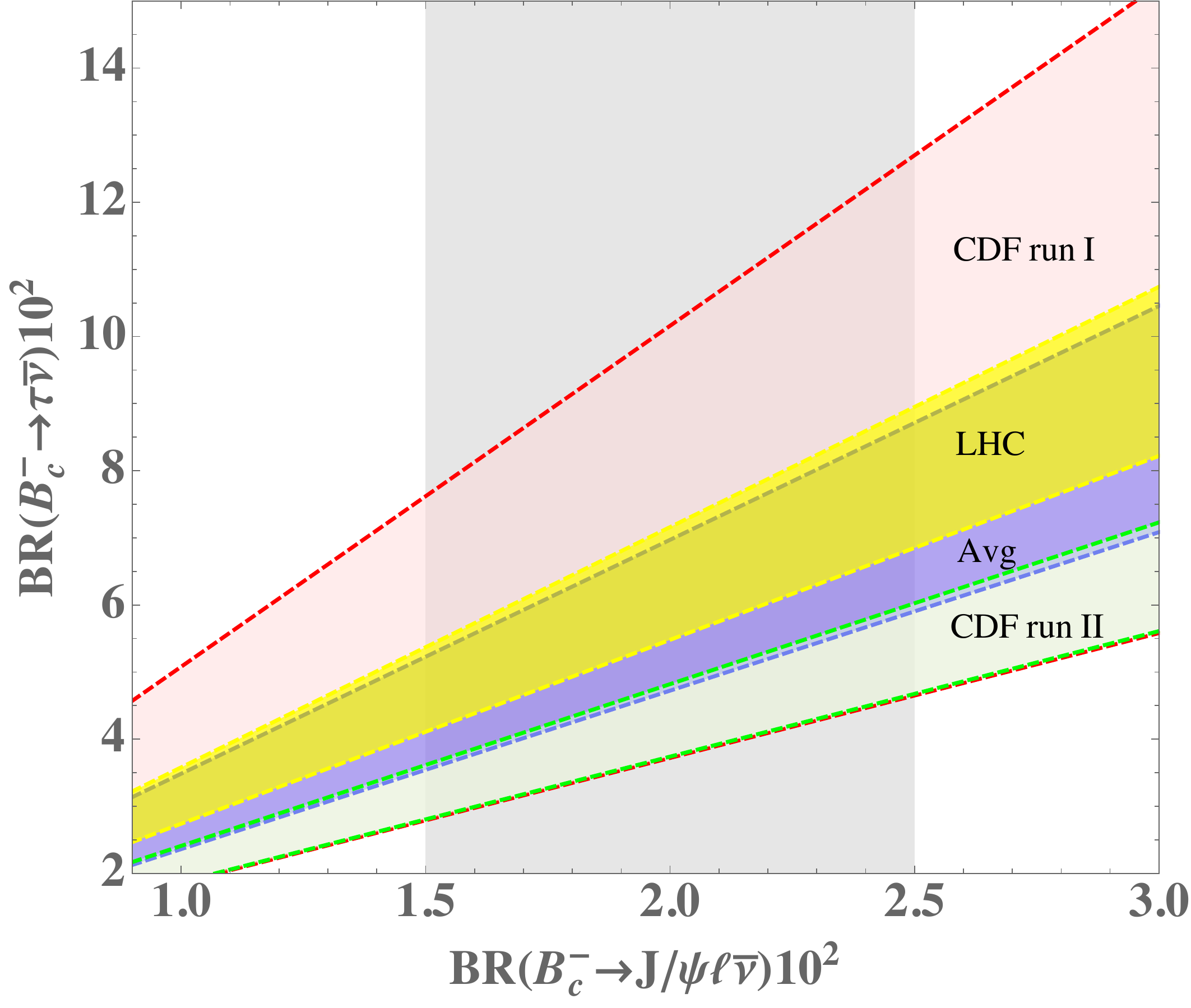}
 \caption{The limit on BR$(B^-_c\to \tau\bar\nu$) as a 
function of BR$(B^-_c \to J/\psi \ell \bar\nu)$ where the different bands
denote the data from CDF Run I (red), Run II (green), LHC (yellow), 
and the average of all three (blue).}
\label{fig:Bc_JPsi}
\end{figure}

We now substitute the expression for $f_c/f_u$  in  
the expression for BR$(B^-_c\to \tau\bar\nu)$ in eq~.(\ref{Bc}).
%Using the world average $B_u=1.06\pm 0.19
%\times 10^{-4}$ \cite{Amhis:2016xyh} in eq.~(\ref{Btaunu_exp}), 
%the L3 bound 
Using ${\rm BR_{eff}}< 5.7\times 10^{-4}$ \cite{Acciarri:1996bv}, and the
Tevatron/LHC data for $f_c/f_u$ in eq.~(\ref{fcfu_tev_lhc}) one obtains the expression:
\begin{equation}
%{\rm BR}(B_c\to \tau\bar\nu)<{\rm BR}(B_c\to J/\psi
%\ell^\pm \nu_{\ell})\left[2.64\pm 0.52\right]\,\,\,\,\,\,{\rm (Tevatron\,data)}\,.
{\rm BR}(B^-_c\to \tau\bar\nu)<{\rm BR}(B^-_c\to J/\psi
\ell \bar \nu)  \left\{ \begin{array}{c}
             2.64\pm 0.52~~{\rm (Tevatron\,data)} \,, \\
             3.16\pm 0.42~~{\rm (LHC\,data)}\,. ~~~~~ \end{array} \right.
\label{Bc_lim_tev}
\end{equation}
%Using the expression for $f_c/f_u$ with LHC data in
% eq.~(\ref{fcfu_lhc}) one has:
%\begin{equation}
%{\rm BR}(B_c\to \tau\bar\nu)<{\rm BR}(B_c\to J/\psi
%\ell^\pm \nu_{\ell})\left[3.16\pm 0.42\right]\,\,\,\,\,\,{\rm (LHC\,data)}\,.
%\label{Bc_lim_lhc}
%\end{equation}
Here the error is from $B_u$, BR$(B^-_u\to J/\psi K^-)_{exp}$ and 
${\cal R}_{\ell}$, which can be seen from the explicit formula:
\begin{equation}
{\rm BR}(B^-_c\to \tau\bar\nu)={\rm BR}(B^-_c\to J/\psi\ell \bar \nu)
\frac{1}{{\cal R}_{\ell}}
\frac{{\rm BR_{eff}}-B^{exp}_u}
{{\rm BR}(B^-_u\to J/\psi K^-)_{exp}}\,.
\end{equation}
Various theoretical calculations for $BR(B^-_c\to J/\psi
\ell \bar \nu)$ are available \cite{Chang:1992pt,Anisimov:1998uk,Kiselev:1999sc,AbdElHady:1999xh,Colangelo:1999zn,Nobes:2000pm,Ebert:2003cn,Ivanov:2005fd,Ivanov:2006ni,Hernandez:2006gt,Wang:2008xt,Ke:2013yka}. In Table~\ref{R_ratio} we present
the bounds on the ratio $R$ defined by
\begin{equation}
R=\frac{{\rm BR}(B^-_c\to \tau\bar\nu)}
{{\rm BR}(B^-_c\to J/\psi\ell \bar\nu)}\,.
\end{equation}
%In Table~\ref{Bc_bound} we present the bounds
%on BR$(B_c\to \tau\nu)$ for values of  BR$(B_c\to J/\psi
%\ell^\pm \nu_{\ell})$ that span the range of the theoretical predictions
%\cite{Chang:1992pt,Anisimov:1998uk,Kiselev:1999sc,AbdElHady:1999xh,Colangelo:19%99zn,Nobes:2000pm,Ebert:2003cn,Ivanov:2005fd,Ivanov:2006ni,Hernandez:2006gt,Wan%g:2008xt,Ke:2013yka}.
%One can see that the weakest bounds (which are obtained for 
%BR$(B_c\to J/\psi
%\ell^\pm \nu_{\ell})=6.7\%$) are still stronger than the bound of
% BR$(B_c\to \tau\nu)< 30\%$ \cite{Alonso:2016oyd} 
%from considering the lifetime of $B_c$.
In Fig.~(\ref{fig:Bc_JPsi}) we show the bounds
on BR$(B^-_c\to \tau\bar\nu)$ as a function of values of BR$(B^-_c\to J/\psi 
\ell \bar\nu)$ that span the range of the theoretical predictions
\cite{Chang:1992pt,Anisimov:1998uk,Kiselev:1999sc,AbdElHady:1999xh,
Colangelo:1999zn,Nobes:2000pm,Ebert:2003cn,Ivanov:2005fd,Ivanov:2006ni,Hernandez:2006gt,Wang:2008xt,Ke:2013yka}.
The four bands are obtained with the measured value of ${\cal R}_\ell$ from  
i) CDF (Run I), ii) CDF Run II, iii) LHC, and iv) the average of all 
three measurements. 
One can see that the weakest bounds 
(which are obtained for 
BR$(B^-_c\to J/\psi
\ell \bar\nu)=2.5\%$) are still stronger than the bound of
 BR$(B^-_c\to \tau\bar\nu)\lesssim 30\%$ \cite{Alonso:2016oyd} 
from considering the lifetime of $B_c$ e.g. with the LHC data alone
one has BR$(B^-_c\to \tau\bar\nu) \lesssim 10\%$. The strongest bounds 
(which are obtained for 
BR$(B^-_c\to J/\psi
\ell  \bar\nu)=1.5\%$) are very close
to the SM prediction of BR$(B^-_c\to \tau\bar\nu) \approx 2\%$ e.g.  with
the CDF run II data alone one has BR$(B^-_c\to \tau\bar\nu) \lesssim 3\%$.

\begin{table}
\center{\begin{tabular}{|l|c|c|c|c|}
\hline
& Tevatron Run I & Tevatron Run II &  LHC & Avg \\\hline
$R$ & $3.47\pm 1.61$ &  $2.14\pm 0.27$ & $3.16\pm 0.42$ & $2.92 \pm 0.56$ \\\hline
\end{tabular} }
\caption{Bound on $R={\rm BR}(B^-_c\to \tau\bar\nu)/
{\rm BR}(B^-_c\to J/\psi\ell \bar\nu)$.}
\label{R_ratio}
\end{table}

%\begin{table}
%\center{\begin{tabular}{|l|c|c|c|}
%\hline
%BR$(B_c\to J/\psi\ell^\pm \nu_{\ell})$  & 1.15\% & 2.4\% & 6.7\% \\\hline
%Tevatron Run I & 3.99 & 8.33 & 23.3\\\hline
%Tevatron Run II & 2.46 &  5.13 & 14.3\\\hline
%LHCb & 3.62 &  7.56 & 21.1\\\hline
%\end{tabular} }
%\caption{Bound on BR$(B_c\to \tau\bar\nu)$ for a given
%value of BR$(B_c\to J/\psi\ell^\pm \nu_{\ell})$.}
%\label{Bc_bound}
%\end{table}

%\subsection{ Description for the theoretical predications on $B_c \to J/\psi \e%ll \bar\nu$}
A sizeable uncertainty in the extraction of the bound on 
BR$(B^-_c\to \tau \bar\nu)$ is the theoretical prediction for 
BR$(B^-_c \to J/\psi \ell \bar\nu)$, of which there are several calculations.  
The estimated values for BR$(B^-_c \to J/\psi \ell \bar\nu)$  
mostly fall within the range 
$1.50-2.50$  $\%$~\cite{Chang:1992pt,Anisimov:1998uk,Kiselev:1999sc,AbdElHady:1999xh,Colangelo:1999zn,Nobes:2000pm,Ebert:2003cn,Ivanov:2005fd,Ivanov:2006ni,Hernandez:2006gt,Wang:2008xt,Ke:2013yka}, 
with the exception being a value of 
$6.7\%$ that was obtained in~\cite{Qiao:2012vt}. 
Without further information from experimental measurements or 
from first-principle QCD calculations, it is not clear which value 
of BR$(B^-_c \to J/\psi \ell \bar\nu)$ to select from the widespread values 
when evaluating the constraint on BR$(B^-_c\to \tau\bar\nu)$. 
Recently, the HPQCD collaboration has made progress in the calculations of 
the form factors for the decays $B^-_c\to J/\psi$~\cite{Colquhoun:2016osw},  
and the obtained (preliminary) results are as follows:
 \begin{align}
% & f_+ = [ a\pm b, a\pm b]\,, \  f_{-}=[a\pm b,\, a \pm b] \nonumber \\
 %
 A_1 = [ 0.49,\, 0.79]\,, \  V=[0.77, \, \text{None}]\,.
 \end{align}
Here $F=[F(q^2=0), F(q^2_{max})]$ denotes the values of a form factor at 
$q^2=0$ and $q^2_{max}$.  We note that all the errors have not been fully
determined, but the total error in the form factors is expected to be of the 
order of $10\%$ or less.
Taking the HPQCD results as a theoretical guidance, we select the QCD model results 
from~\cite{Chang:1992pt,Anisimov:1998uk,Kiselev:1999sc,AbdElHady:1999xh,Colangelo:1999zn,Nobes:2000pm,Ebert:2003cn,Ivanov:2005fd,Ivanov:2006ni,Hernandez:2006gt,Wang:2008xt,Ke:2013yka,Qiao:2012vt} 
for which the predicted form factors at $q^2=0$  are within $15\%$ of the values of the HPQCD calculation. 
Accordingly, the results of the selected  QCD approaches are shown in Table~\ref{tab:FFs}, where the 
last column is the predicted BR($B^-_c\to J/\psi \ell \bar\nu$). 
It can be clearly seen that values of BR$(B^-_c \to J/\psi \ell \bar \nu)$ in range $\approx (2.0\pm 0.5)\%$ are 
favoured when using the values of the form factors from lattice QCD as a guide.

%\begin{align}
%\langle P' (p_2) | V_\mu| P(p_1) \rangle & = f_{+}(q^2) \left( P_\mu -\frac{P\cdot q}{q^2} q_\mu \right) + f_{-}(q^2) \frac{P\cdot q}%{q^2} q_\mu\,, \\
%
%\langle V (p_2) | V_\mu -A_\mu | P (p_1) \rangle  &=i \frac{V(q^2) \varepsilon_{\mu\nu\rho\sigma}}{m_P + m_V} \epsilon^{*\nu}_V P^\rho q^\sigma -(m_P + m_V)  A_1(q^2) \left( \epsilon^*_{V\mu} - \frac{\epsilon^*_V \cdot q}{q^2} q_\mu \right)  \nonumber \\
%&  + \frac{\epsilon^*_V \cdot q  }{m_P+m_V} A_2(q^2) \left( P_\mu - \frac{P\cdot}{q^2} q_\mu\right) -2 m_P \frac{\epsilon^*\cdot q }{q^2} q_\mu A_0(q^2)\,,
%\end{align}
%where $V_{\mu}(A_\mu)$ denotes the weak (axial) vector current, $\epsilon_V$ is the polarization vector of a vector-meson, %and $P(q)= p_1 \pm p_2$. 

\begin{table}
\caption{Form factors for $B^-_c \to J/\psi$ at $q^2=0$ and $q^2_{max}$.} \label{tab:FFs}
\begin{center}
\begin{tabular}{|c|c|c|c|}  \hline 
$(F(0), F(q^2_{max}))$  &   $A_1$ & $V$  & BR$(B_c \to J/\Psi \ell \bar \nu)$\\ \hline
HPQCD\cite{Colquhoun:2016osw} & ~~ ( 0.49, 0.79)~~  &~~(0.77, None)~~ & None \\ \hline
NW\cite{Nobes:2000pm} & ~~ (0.53, 0.76\mytabnote)~ & ~(0.73, 1.29$^a$)~~& $1.47\%$ \\ \hline
IKS\cite{Ivanov:2005fd} & ~~ (0.55, 0.85)~~ & ~~(0.83, 1.53)~~ & $2.17\%$ \\ \hline
WSL\cite{Wang:2008xt} &~~ (0.50, 0.80)~~ & ~~(0.74, 1.45)~~ & $1.49\%$ \\ \hline
\end{tabular} 
\end{center}
\end{table}

\section{Impact on $H^\pm$ interpretation of $R(D)$, $R(D^*)$
anomaly}

%A neutral scalar has been discovered at the Large Hadron Collider (LHC), and its properties are consistent with those of the
% Higgs boson of the Standard Model (SM). Charged Higgs bosons $H^{\pm}$ are predicted in many well-motivated extensions of 
%the SM e.g. models which contain two or more $SU(2)XU(1)$ scalar doublets (which includes supersymmetric models).
% Such particles are being searched for directly at the LHC, and their non-observance leads to excluded regions of 
%the parameters i) $\tan\beta$ and ii) charged Higgs mass ($m_{H^\pm}$). 
The following ratios $R(D)$ and $R(D^*)$ are defined as follows:
\begin{equation}
R(D)=\frac{{\rm BR}(B\to D\tau\nu)}{{\rm BR}(B\to D\ell\nu)};\;\;\;\;\;R(D^*)=\frac{{\rm BR}(B\to D^*\tau\nu)}{{\rm BR}(B\to D^*\ell\nu)}\,.
\end{equation}
The current world averages \cite{HFLAV} of their measurements at BABAR
\cite{Lees:2012xj,Lees:2013uzd}, BELLE 
\cite{Huschle:2015rga,Sato:2016svk,{Hirose:2016wfn}}
and LHCb \cite{Aaij:2015yra} are:
\begin{equation}
R(D)=0.407\pm 0.039\pm 0.024  ;\;\;\;\;\;R(D^*)=0.304\pm 0.013\pm 0.007\,.
\end{equation}
The predictions in the SM for $R(D)$ \cite{Lattice:2015rga,Na:2015kha}
and $R(D^*)$ \cite{Fajfer:2012vx} are given by:
\begin{equation}
R(D)=0.300\pm 0.008  ;\;\;\;\;\;R(D^*)=0.252\pm 0.003\,.
\label{RD_pred}
\end{equation}
The above measurements of $R(D)$ and $R(D^*)$ exceed the SM predictions by $2.3\sigma$ and $3.4\sigma$ respectively. 
Taking into account the $R(D)$-$R(D^*)$ correlation, the deviation with respect to the SM prediction is $4.1\sigma$. Consequently,
there have been many works that explain this deviation by invoking the contribution of new physics particles. One such
candidate particle is $H^\pm$, 
which is predicted in many well-motivated extensions of 
the SM e.g. models that contain two or more $SU(2)\otimes U(1)$ scalar doublets (which includes supersymmetric models).

It has been shown that an $H^\pm$ from a Two Higgs Doublet Model (2HDM) with type II couplings and natural flavour conservation
cannot accommodate the above data for $R(D)$ and $R(D^*)$. However, an $H^\pm$ in a 2HDM without natural flavour conservation
(called the ``generic 2HDM'' or ``Type III 2HDM'', in which both Higgs doublets couple to each fermion type) can give rise to the
measured values of $R(D)$ and $R(D^*)$ 
\cite{Crivellin:2012ye,Celis:2012dk,Crivellin:2013wna,Celis:2016azn,
Crivellin:2015mga,Chen:2017eby,Arbey:2017gmh}.

However, recently it has been shown that there is a correlation 
between $R(D^*)$ and BR$(B^-_c\to \tau\bar\nu)$, and any enhancement
of the former by $H^\pm $ gives rise to an enhancement of the latter \cite{Alonso:2016oyd}. In \cite{Alonso:2016oyd} the direct
limit on BR$(B^-_c\to \tau\bar\nu)$ (that is derived in section II) is not considered. Instead, an indirect limit
of BR$(B^-_c\to \tau\bar\nu)\lesssim 30\%$ was derived by considering the current measurement of the lifetime of $B_c$ i.e. the partial decay width 
of $B^-_c\to \tau\bar\nu$ is bounded from the knowledge of the total decay width (inverse of lifetime) of $B_c$. 
The bound BR$(B^-_c\to \tau\bar\nu)\lesssim 30\%$ restricts $R(D^*)$ to values 
$\lesssim 0.275$, which at the moment slightly
disfavours an explanation of the $R(D)$ and $R(D^*)$ anomaly from $H^\pm$ alone. The bound  BR$(B^-_c\to \tau\bar\nu)\lesssim 30\%$
has been implemented in subsequent studies that consider $H^\pm$ as a candidate for explaining the  $R(D)$ and $R(D^*)$ anomaly e.g. \cite{Celis:2016azn}.

We now study the effect of $H^\pm$ on $R(D)$, $R(D^*)$ and 
$B^-_c\to \tau\bar\nu$.
In $R(D)$ and $R(D^*)$ the underlying quark decay is $b\to c \tau \bar \nu$,
while $B^-_c\to \tau\bar\nu$ proceeds via annihilation of the meson to
a $W^\pm$ or $H^\pm$. The effective Lagrangian for the contribution of 
$W^\pm$ and $H^\pm$ bosons to all three decays is given by: 
 \begin{align}
 {\cal H}_{\rm eff} & = \frac{ G_F V_{cb}}{\sqrt{2}}  \left[ (\bar c b)_{V-A} (\bar\tau \nu_\tau)_{V-A} + \left(C^\tau_{R} (\bar c b)_{S+P}
 + C^\tau_{L} (\bar c b)_{S-P} \right) (\bar\tau \nu_\tau)_{S-P}\right]\,,
 \end{align}
where $(\bar {f'} f)_{V-A}=\bar{f'} \gamma_\mu (1- \gamma_5) f$, $(\bar {f'} f)_{S\pm P}=\bar{f'}  (1\pm \gamma_5) f$, and $C^\tau_{L,R}$ are the effective couplings which combine the quark and tau-lepton Yukawa couplings. In general the neutrino can be any flavour,
but since the enhancement of 
$R(D^{(*)})$ is mainly from the constructive interference of $H^\pm$ 
with the SM contribution, we only consider $\nu_\tau$ in the effective Lagrangian. 
The couplings $C^\tau_L$ and $C^\tau_R$
are functions of $\tan\beta$ and $m_{H^\pm}$ in a 2HDM with natural flavour conservation. In a generic 2HDM,
$C^\tau_L$ and $C^\tau_R$ have an additional dependence on parameters that lead to flavour changing neutral currents see e.g. \cite{Chen:2017eby}.

\begin{figure}[phtb]
\includegraphics[width=85mm]{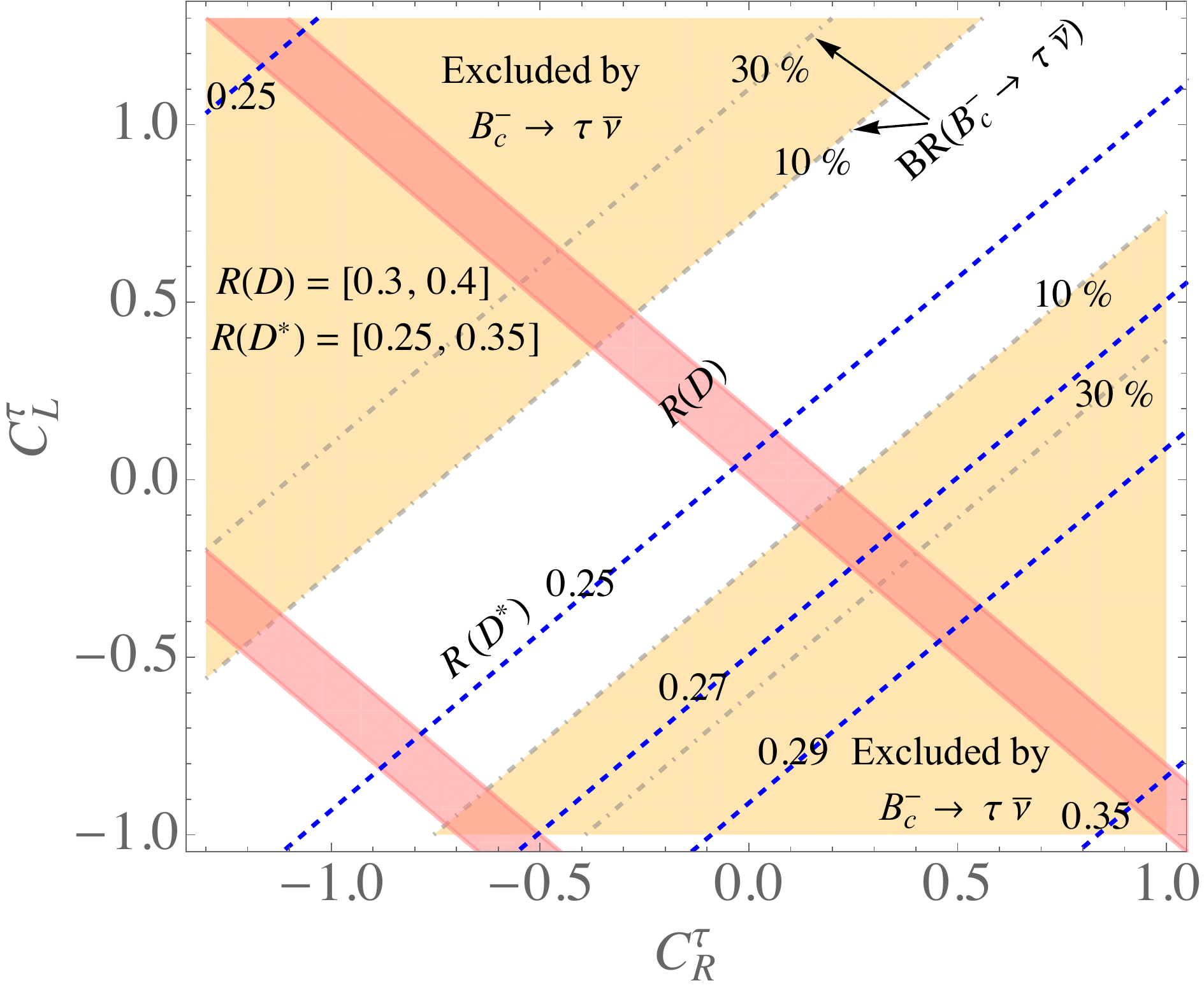}
 \caption{ Contours for $R(D)$ (band), $R(D^*)$, and BR$(B^-_c \to \tau \bar\nu)$ as a function of $C^\tau_{R,L}$, where the ranges of $R(D)=[0.30,0.4]$ and $R(D^*)=[0.25,0.35]$, and BR$(B^-_c\to \tau \bar\nu)\lesssim 30\%, 10\%$ are taken. }
\label{fig:RD_RDv_Bc}
\end{figure}

To demonstrate the impact of 
$B^-_c\to \tau \bar \nu$ on $R(D^{(*)})$, we show the contours for $R(D)$ (band), $R(D^*)$ (dashed), and BR$(B^-_c \to \tau \bar \nu)$ (dash-dotted) as a function of $C^\tau_R$ and $C^\tau_L$ in Fig.~(\ref{fig:RD_RDv_Bc}), where  the estimations for $R(D)$ and $R(D^*)$ are based on the formulae in~\cite{Fajfer:2012vx};  the ranges of  $R(D)=[0.3, 0.4]$ and $R(D^*)=[0.25,0.35]$, and BR$(B^-_c\to \tau \bar \nu)< 30\%, 10\%$ are used. It can be seen that the bound 
BR$(B^-_c\to \tau \bar \nu)<10\%$ reduces the maximum allowed
value of $R(D^*)$ to $\sim 0.26$. Hence in context of an 
enhancement of $R(D)$ by $H^\pm$ alone, 
the maximum allowed value of $R(D^*)$ is reduced
from  $R(D^*)\sim 0.275$ (for BR$(B^-_c\to \tau \bar \nu)\lesssim 
30\%$ and see e.g. 
\cite{Celis:2016azn}) to $R(D^*)\sim 0.26$ i.e. to within $3\sigma$ of
the SM prediction for  $R(D^*)$ in eq.~(\ref{RD_pred}).
We note that other models with new physics particles 
(e.g. leptoquarks) can give rise to other terms
in the effective Hamiltonian for the  $cb\tau\nu$ vertex. These models
are not strongly constrained by BR$(B^-_c\to \tau \bar \nu)$, as discussed 
in \cite{Alonso:2016oyd}.

\begin{figure}[phtb]
\includegraphics[width=85mm]{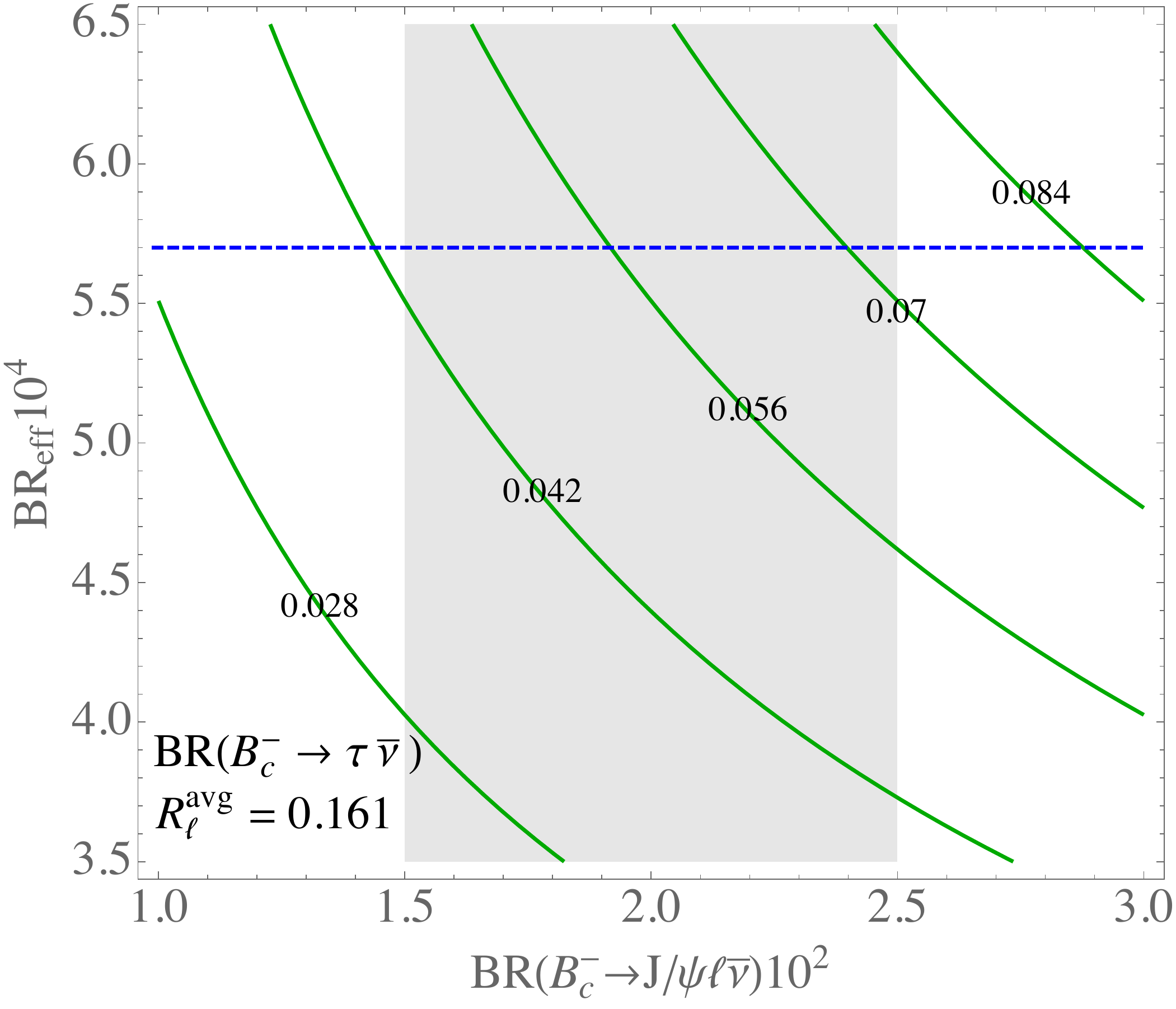}
 \caption{Contours for BR$(B^-_c\to \tau \bar\nu)$ as a function of
BR$(B^-_c\to J/\psi \ell \bar \nu$) and  ${\rm BR_{eff}}$, for ${\cal R}_\ell=0.161$. 
The current limit ${\rm BR_{eff}<5.7\times 10^{-4}}$ is shown.}
\label{fig:BReff}
\end{figure}

Prospects for more precise measurements of  $R(D)$ and $R(D^*)$ are good. 
Although LHCb has currently only measured  
$R(D^*)$ (for two separate decay modes of the $\tau$, and with the data
taken at $\sqrt s=7$ TeV and $\sqrt s=8$ TeV) 
it is capable of measuring $R(D)$ \cite{Bozzi:2017suf}. 
Measurements with data taken at $\sqrt s=13$ TeV data will further reduce the
error in the world averages of both observables. 
The BELLE-II experiment will eventually have roughly fifty times as much integrated luminosity 
as the final integrated luminosities from the $B$ factories (BABAR and BELLE), 
and hence significantly more precise measurements of $R(D)$ and $R(D^*)$
will become available.  
In contrast, it is challenging for the LHC experiments 
to directly measure (or set direct limits on) BR($B^-_c\to \tau\bar\nu$). 
As discussed in \cite{Akeroyd:2008ac}, the best prospect for observing the decay $B_c\to \tau\bar\nu$
is a period of operation of an $e^+e^-$ linear collider at $\sqrt s\sim 91$ GeV. We note that 
the L3 limit \cite{Acciarri:1996bv} only used $40\%$ of the available data taken at $\sqrt s\sim 91$ GeV.
If the full L3 data sample were used, the limit ${\rm BR_{eff}<5.7\times 10^{-4}}$ could be improved, or
even evidence for first observation of $B^-_c\to \tau\bar\nu$ could be obtained.
As shown in Fig.~(\ref{fig:Bc_JPsi}), the strongest bound on
BR$(B^-_c\to \tau\bar\nu)$ is $\lesssim 3\%$, 
which is just above the SM prediction 
of $\sim 2\%$.
%%%%
In Fig~.(\ref{fig:BReff}) 
contours for BR$(B^-_c\to \tau \bar\nu)$ are shown as a function of
BR$(B^-_c\to J/\psi \ell \bar\nu$) and ${\rm BR_{eff}}$.
We take ${\cal R}_\ell=0.161$, which is the central value of the average
of the CDF Run I, CDF Run II and LHC measurements. The shaded region 
corresponds to the range
of theoretical predictions of BR$(B^-_c\to J/\psi \ell \bar \nu)$.
It was suggested in \cite{Akeroyd:2008ac} that sensitivity to
${\rm BR_{eff}}\sim 4\times 10^{-4}$ might be reached if L3 used 
all the data taken at $\sqrt s\sim 91$ GeV. 
From  Fig~.(\ref{fig:BReff}) it can be seen that this limit is close
to the value of ${\rm BR_{eff}}$ that is obtained for a SM-like value 
$(\approx 2\%)$ for
BR$(B^-_c\to \tau \bar\nu)$.

%%%%%%%%%%%%%%%%%%%%%%%%%%%%%%%%%%%%%%%%%%%%%%%%%%%%%

\section{Conclusions}  
As discussed in \cite{Mangano:1997md,Akeroyd:2008ac},
LEP data taken at the $Z$ peak constrained a combination of 
the decays $B^-_u \to \tau \bar \nu$ and $B^-_c \to \tau \bar \nu$. 
This is the only data that directly constrains the magnitude of
BR($B^-_c \to \tau \bar \nu$).
From the L3 limit \cite{Acciarri:1996bv} 
we derived for the first time an explicit 
bound on BR$(B^-_c \to \tau \bar \nu)$.  The bound can be
conveniently written in terms of experimentally
determined quantities and just one theoretical input parameter, 
which is the branching ratio of $B^-_c\to J/\psi \ell \bar\nu$.
Using the theoretically preferred range for BR($B^-_c\to J/\psi \ell \bar\nu$)
we showed that BR$(B^-_c \to \tau \bar \nu)\lesssim 10\%$, which
is considerably stronger than the bound 
from considering the lifetime of $B^-_c$ \cite{Alonso:2016oyd}. 

It is known that any bound on BR$(B^-_c \to \tau \bar \nu)$
has consequences for an explanation of the 
$R(D)$ and $R(D^*)$ anomaly in terms of an $H^\pm$ alone. In such
scenarios, any enhancement of $R(D^*)$ leads to an enhancement of 
BR$(B^-_c \to \tau \bar \nu)$.
Our new bound on BR$(B^-_c \to \tau \bar \nu)$
further reduces the maximum enhancement of $R(D^*)$ from an $H^\pm$.
Thus if future values of $R(D)$ stay significantly higher than the SM 
predictions, any explanation that uses $H^\pm$ alone would require 
the measured value of $R(D^*)$ to approach values that are closer to the SM prediction. 

The observables $R(D)$, $R(D^*)$ and $B^-_c \to \tau \bar \nu$ 
all proceed via the same effective Lagrangian, and
thus measurement of BR($B^-_c \to \tau \bar \nu$) would provide
independent information on the relevant couplings. 
Direct searches for $B^-_c \to \tau \bar \nu$ 
at the LHC are challenging. However, as stressed in \cite{Akeroyd:2008ac},
 a stronger limit on BR$(B^-_c \to \tau \bar \nu)$
(or even first observation of this decay) could be obtained if
the L3 collaboration used all their data to update the limit
in  \cite{Acciarri:1996bv} (which used $\sim 40\%$ of the available data).
Operation of an $e^+e^-$ linear collider at the $Z$ peak would have 
sensitivity to the SM branching ratio of $B^-_c \to \tau \bar \nu$.

\section*{Acknowledgements}

This work was partially supported by the Ministry of 
Science and Technology of Taiwan,  
under grant MOST-106-2112-M-006-010-MY2 (CHC). We thank A. Lytle for
useful comments.

\end{document}